\documentstyle[12pt,epsf]{article}

\catcode`\@=11
\long\def\@makefntext#1{
\protect\noindent \hbox to 3.2pt {\hskip-.9pt
$^{{\ninerm\@thefnmark}}$\hfil}#1\hfill}                

\def\@makefnmark{\hbox to 0pt{$^{\@thefnmark}$\hss}}  

\def\ps@myheadings{\let\@mkboth\@gobbletwo
\def\@oddhead{\hbox{}
\rightmark\hfil\ninerm\thepage}
\def\@oddfoot{}\def\@evenhead{\ninerm\thepage\hfil
\leftmark\hbox{}}\def\@evenfoot{}
\def\sectionmark##1{}\def\subsectionmark##1{}}

\renewcommand{\thefootnote}{\fnsymbol{footnote}}

\newcounter{sectionc}\newcounter{subsectionc}\newcounter{subsubsectionc}
\renewcommand{\section}[1] {\vspace*{0.6cm}\addtocounter{sectionc}{1}
\setcounter{subsectionc}{0}\setcounter{subsubsectionc}{0}\noindent
        {\normalsize\bf\thesectionc. #1}\par\vspace*{0.4cm}}
\renewcommand{\subsection}[1] {\vspace*{0.6cm}\addtocounter{subsectionc}{1}
        \setcounter{subsubsectionc}{0}\noindent
        {\normalsize\it\thesectionc.\thesubsectionc. #1}\par\vspace*{0.4cm}}
\renewcommand{\subsubsection}[1]
{\vspace*{0.6cm}\addtocounter{subsubsectionc}{1}
        \noindent
{\normalsize\rm\thesectionc.\thesubsectionc.\thesubsubsectionc.
        #1}\par\vspace*{0.4cm}}

\newcounter{appendixc}
\newcounter{subappendixc}[appendixc]
\newcounter{subsubappendixc}[subappendixc]

\renewcommand{\appendix}[1] {\vspace*{0.6cm}
        \refstepcounter{appendixc}
        \setcounter{figure}{0}
        \setcounter{table}{0}
        \setcounter{equation}{0}
        \renewcommand{\thefigure}{\Alph{appendixc}.\arabic{figure}}
        \renewcommand{\thetable}{\Alph{appendixc}.\arabic{table}}
        \renewcommand{\theappendixc}{\Alph{appendixc}}
        \renewcommand{\theequation}{\Alph{appendixc}.\arabic{equation}}
        \noindent{\bf Appendix \theappendixc #1}\par\vspace*{0.4cm}}

\def\abstracts#1{{

\centering{\begin{minipage}{12.2truecm}\footnotesize\baselineskip=12pt\noindent
        \centerline{\footnotesize ABSTRACT}\vspace*{0.3cm}
        \parindent=0pt #1
        \end{minipage}}\par}}

\renewenvironment{thebibliography}[1]
        {\begin{list}{\arabic{enumi}.}
        {\usecounter{enumi}\setlength{\parsep}{0pt}
\setlength{\leftmargin 1.25cm}{\rightmargin 0pt}
         \setlength{\itemsep}{0pt} \settowidth
        {\labelwidth}{#1.}\sloppy}}{\end{list}}

\newcounter{itemlistc}
\newcounter{romanlistc}
\newcounter{alphlistc}
\newcounter{arabiclistc}

\newcommand{\fcaption}[1]{
        \refstepcounter{figure}
        \setbox\@tempboxa = \hbox{\footnotesize Fig.~\thefigure. #1}
        \ifdim \wd\@tempboxa > 6in
           {\begin{center}
        \parbox{6in}{\footnotesize\baselineskip=12pt Fig.~\thefigure. #1}
            \end{center}}
        \else
             {\begin{center}
             {\footnotesize Fig.~\thefigure. #1}
              \end{center}}
        \fi}

\newcommand{\tcaption}[1]{
        \refstepcounter{table}
        \setbox\@tempboxa = \hbox{\footnotesize Table~\thetable. #1}
        \ifdim \wd\@tempboxa > 6in
           {\begin{center}
        \parbox{6in}{\footnotesize\baselineskip=12pt Table~\thetable. #1}
            \end{center}}
        \else
             {\begin{center}
             {\footnotesize Table~\thetable. #1}
              \end{center}}
        \fi}

\def\@citex[#1]#2{\if@filesw\immediate\write\@auxout
        {\string\citation{#2}}\fi
\def\@citea{}\@cite{\@for\@citeb:=#2\do
        {\@citea\def\@citea{,}\@ifundefined
        {b@\@citeb}{{\bf ?}\@warning
        {Citation `\@citeb' on page \thepage \space undefined}}
        {\csname b@\@citeb\endcsname}}}{#1}}

\newif\if@cghi
\def\cite{\@cghitrue\@ifnextchar [{\@tempswatrue
        \@citex}{\@tempswafalse\@citex[]}}
\def\citelow{\@cghifalse\@ifnextchar [{\@tempswatrue
        \@citex}{\@tempswafalse\@citex[]}}
\def\@cite#1#2{{$\null^{#1}$\if@tempswa\typeout
        {IJCGA warning: optional citation argument
        ignored: `#2'} \fi}}

 1
 1
 1

\font\ninerm=cmr9

\newcommand{\be}{\begin{equation}}
\newcommand{\ee}{\end{equation}}
\newcommand{\ba}{\begin{eqnarray}}
\newcommand{\ea}{\end{eqnarray}}
\newcommand{\ds}{\displaystyle}
\newcommand{\baa}{\begin{eqnarray*}}
\newcommand{\eaa}{\end{eqnarray*}}
\newcommand{\bb}{\begin{thebibliography}}
\newcommand{\eb}{\end{thebibliography}}
\newcommand{\ci}[1]{\cite{#1}}
\newcommand{\bi}[1]{\bibitem{#1}}
\newcommand{\lab}[1]{\label{#1}}
\newcommand{\re}[1]{(\ref{#1})}

\newcommand{\bit}{\begin{itemize}}
\newcommand{\eit}{\end{itemize}}
\newcommand{\al}{\alpha}
\newcommand{\bt}{\beta}

\newcommand{\s}{\sigma}

\newcommand{\r}{\rho}

\newcommand{\gag}{$\gamma^*\gamma^*\rightarrow\pi^\circ$}
\newcommand{\fgg}{\mbox{$F_{\gamma^*\gamma^*\rightarrow\pi^\circ}$}}

\begin{document}

\centerline{\normalsize\bf FORM FACTOR OF THE PROCESS \gag\ FOR SMALL}
\baselineskip=16pt
\centerline{\normalsize\bf VIRTUALITY OF ONE OF THE PHOTONS AND QCD SUM RULES}

\centerline{\footnotesize A.V.RADYUSHKIN}
\baselineskip=13pt
\centerline{\footnotesize\it Old Dominion University, Norfolk, VA 23529, USA;}
\baselineskip=12pt
\centerline{\footnotesize\it CEBAF,
 Newport News, VA 23606, USA}
\centerline{\footnotesize E-mail: radyush@cebaf.gov}
\vspace*{0.3cm}
\centerline{\footnotesize and}
\vspace*{0.3cm}
\centerline{\footnotesize R.RUSKOV\footnote{Contribution to the
PHOTON95 conference, Sheffield (1995)}
}
\baselineskip=13pt
\centerline{\footnotesize\it Laboratory of Theoretical Physics, JINR, Dubna,
Russia}
\centerline{\footnotesize E-mail: ruskovr@thsun1.jinr.dubna.su}

\vspace*{0.9cm}
\abstracts{We extend the QCD sum rule analysis of the  \fgg$(q_1^2,q_2^2)$
form factor  into the region where one of the photons
has small virtuality:
$|q_1^2|\ll|q_2^2|\geq 1{\mbox{ GeV}}^2$.
In this kinematics,  one should perform  an additional
factorization
of short- and long-distance contributions.
The extra  long-distance sensitivity  of the
three-point amplitude is described by two-point
correlators (bilocals), and the low-momentum dependence
of the correlators involving composite operators
of two lowest twists  is extracted
 from auxiliary QCD sum rules.
Our estimates  for \fgg$(q_1^2=0,q_2^2)$ are in good
agreement with  existing  experimental data.}

\normalsize\baselineskip=15pt
\setcounter{footnote}{0}
\renewcommand{\thefootnote}{\alph{footnote}}

\section{Introductory Remarks}

In this work we present the result of our calculation \ci{one,RadRu94} of the
\fgg$(q_1^2,q_2^2)$ form factor at small virtualities of one of the photons
$|q_1^2|\ll{|q_2^2|}\geq{1}\mbox{ GeV}^2$
within the QCD sum rule method \ci{SVZ79}. In our first paper \ci{one}, we
formulated a  QCD sum rule
approach  to the problem and analyzed  the structure of the OPE
for the relevant three-point correlation function
\be
{\cal{F}}_{\al\mu\nu}(q_1,q_2)={\it{i}}\int  e^{-iq_{1}x - iq_{2}y} \,
\langle 0 |T\left\{J_{\mu}(x)\,J_{\nu}(y)\,j_{\al}^5(0)\right\}| 0 \rangle
  d^4x\,d^4y\
\lab{eq:corr},
\ee
with a particular emphasis on modifications
required by the presence of the infrared (mass) singularities
specific for  the  $q_1^2\rightarrow 0$ limit.
On Fig.1 we illustrate
the  schematic structure
of the modified OPE for the three-point
function Eq.\re{eq:corr}.

The first row in   Fig.1 corresponds to
the usual operator expansion for the three-point correlation function.
It was constructed \ci{one} using the standard approach
 \ci{IoSm82,NeRa82}   valid in the kinematics when all
the momentum invariants are large.
In the  $q_1^2 \to 0$ limit, this OPE is
singular.  In  particular, the  condensate terms
  explode like inverse powers of
 $q_1^2$. The perturbative term
(the triangle loop) is formally finite in the
small-$q_1^2$ limit, but it contains non-analytic contributions like
$q_1^2\ln{(-q_1^2)}$ and $q_1^4\ln{(-q_1^2)}$.
Just like in the pion FF case \cite{NeRa84b},
because of these mass singularities,
one should  perform an additional  factorization of  short and long
distance contributions (see also
refs.\ci{BalY83}-\ci{BeiNeRa88}).

%


\unitlength=2.7pt

\special{em:linewidth 0.4pt}

\linethickness{0.4pt}

\begin{picture}(163.00,42.40)

\put(22.95,21.37){\circle{14.00}}

\put(19.96,33.42){\oval(1.96,1.96)[l]}

\put(19.96,35.42){\oval(1.96,2.04)[r]}

\put(19.96,37.42){\oval(1.96,1.96)[l]}

\put(19.96,39.42){\oval(1.96,2.04)[r]}

\put(19.96,41.42){\oval(1.96,1.96)[l]}

\put(10.98,14.35){\oval(1.96,2.04)[b]}

\put(8.98,14.35){\oval(2.04,2.04)[t]}

\put(6.98,14.35){\oval(1.96,2.04)[b]}

\put(4.98,14.35){\oval(2.04,2.04)[t]}

\put(2.98,14.35){\oval(1.96,2.04)[b]}

\put(16.94,18.33){\line(1,1){8.97}}

\put(27.95,26.33){\line(-1,-1){10.04}}

\put(19.96,15.40){\line(1,1){8.97}}

\put(29.91,22.33){\line(-1,-1){8.00}}

\put(15.96,20.37){\line(1,1){8.00}}

\put(21.94,28.40){\line(-1,-1){6.00}}

\put(15.93,21.37){\line(1,1){7.03}}

\put(24.94,28.40){\line(-1,-1){9.01}}

\put(15.93,18.37){\line(1,1){10.03}}

\put(26.92,27.38){\line(-1,-1){9.96}}

\put(16.96,16.39){\line(1,1){10.98}}

\put(28.97,26.35){\line(-1,-1){10.98}}

\put(18.93,15.37){\line(1,1){10.03}}

\put(29.92,24.38){\line(-1,-1){9.96}}

\put(29.92,23.35){\line(-1,-1){8.93}}

\put(22.96,14.34){\line(1,1){6.96}}

\put(29.92,20.35){\line(-1,-1){6.00}}

\put(11.98,14.34){\line(6,1){6.96}}

\put(15.93,20.35){\line(-2,-3){4.03}}

\put(19.96,32.36){\line(-1,-3){1.98}}

\put(24.94,28.40){\line(-5,4){4.98}}

\put(14.03,40.29){\makebox(0,0)[cc]{$q_1^2$}}

\put(1.01,18.31){\makebox(0,0)[cc]{$q_2^2$}}

\put(13.03,11.38){\makebox(0,0)[cc]{y}}

\put(24.01,33.35){\makebox(0,0)[cc]{x}}

\put(34.96,23.35){\makebox(0,0)[cc]{0}}

\put(38.02,16.37){\makebox(0,0)[cc]{$p^2$}}

\put(33.04,19.29){\makebox(0,0)[cc]{$\times$}}

\put(21.00,28.31){\line(-1,-1){4.99}}

\put(30.02,19.36){\line(-1,-1){4.99}}

\put(48.00,23.33){\makebox(0,0)[cc]{$=$}}

\put(163.00,23.33){\makebox(0,0)[cc]{$+\cdots$}}

\put(33.00,19.32){\line(-5,-2){4.99}}

\put(33.00,19.32){\line(-1,1){3.98}}

\put(107.00,23.33){\makebox(0,0)[cc]{$+$}}

\put(76.95,33.09){\oval(1.96,1.96)[l]}

\put(76.95,35.09){\oval(1.96,2.04)[r]}

\put(76.95,37.09){\oval(1.96,1.96)[l]}

\put(76.95,39.09){\oval(1.96,2.04)[r]}

\put(76.95,41.09){\oval(1.96,1.96)[l]}

\put(67.97,14.02){\oval(1.96,2.04)[b]}

\put(65.97,14.02){\oval(2.04,2.04)[t]}

\put(63.97,14.02){\oval(1.96,2.04)[b]}

\put(61.97,14.02){\oval(2.04,2.04)[t]}

\put(59.97,14.02){\oval(1.96,2.04)[b]}

\put(71.02,39.96){\makebox(0,0)[cc]{$q_1^2$}}

\put(58.00,17.98){\makebox(0,0)[cc]{$q_2^2$}}

\put(95.01,16.04){\makebox(0,0)[cc]{$p^2$}}

\put(90.03,18.96){\makebox(0,0)[cc]{$\times$}}

\put(68.99,14.00){\line(4,1){21.00}}

\put(89.99,19.33){\line(-1,1){13.00}}

\put(76.99,32.33){\line(-2,-5){7.33}}

\put(130.95,33.09){\oval(1.96,1.96)[l]}

\put(130.95,35.09){\oval(1.96,2.04)[r]}

\put(130.95,37.09){\oval(1.96,1.96)[l]}

\put(130.95,39.09){\oval(1.96,2.04)[r]}

\put(130.95,41.09){\oval(1.96,1.96)[l]}

\put(121.97,14.02){\oval(1.96,2.04)[b]}

\put(119.97,14.02){\oval(2.04,2.04)[t]}

\put(117.97,14.02){\oval(1.96,2.04)[b]}

\put(115.97,14.02){\oval(2.04,2.04)[t]}

\put(113.97,14.02){\oval(1.96,2.04)[b]}

\put(125.02,39.96){\makebox(0,0)[cc]{$q_1^2$}}

\put(112.00,17.98){\makebox(0,0)[cc]{$q_2^2$}}

\put(149.01,16.04){\makebox(0,0)[cc]{$p^2$}}

\put(144.03,18.96){\makebox(0,0)[cc]{$\times$}}

\put(122.99,14.00){\line(4,1){21.00}}

\put(143.99,19.33){\line(-1,1){13.00}}

\put(130.99,32.33){\line(-2,-5){7.33}}

\put(136.99,26.00){\line(1,0){2.00}}

\put(140.99,26.00){\line(1,0){2.00}}

\put(144.99,26.00){\line(1,0){2.00}}

\put(116.99,24.00){\line(1,0){2.00}}

\put(120.99,24.00){\line(1,0){2.00}}

\put(124.99,24.00){\line(1,0){2.00}}

\put(23.00,3.00){\makebox(0,0)[cc]{a)}}

\put(79.00,3.00){\makebox(0,0)[cc]{b)}}

\put(134.00,3.00){\makebox(0,0)[cc]{c)}}

\end{picture}

\vspace{0.1cm}


\unitlength=2.50pt

\special{em:linewidth 0.4pt}

\linethickness{0.4pt}

\begin{picture}(158.00,52.78)

\put(13.97,13.73){\oval(1.96,2.04)[b]}

\put(11.97,13.73){\oval(2.04,2.04)[t]}

\put(9.97,13.73){\oval(1.96,2.04)[b]}

\put(7.97,13.73){\oval(2.04,2.04)[t]}

\put(5.97,13.73){\oval(1.96,2.04)[b]}

\put(4.00,17.69){\makebox(0,0)[cc]{$q_2^2$}}

\put(14.99,13.71){\line(1,0){21.00}}

\put(35.99,13.71){\line(-4,5){4.00}}

\put(14.99,13.71){\line(3,5){3.00}}

\put(24.95,41.80){\oval(1.96,1.96)[l]}

\put(24.95,43.80){\oval(1.96,2.04)[r]}

\put(24.95,45.80){\oval(1.96,1.96)[l]}

\put(24.95,47.80){\oval(1.96,2.04)[r]}

\put(24.95,49.80){\oval(1.96,1.96)[l]}

\put(19.02,49.67){\makebox(0,0)[cc]{$q_1^2$}}

\put(24.99,21.71){\makebox(0,0)[cc]{$\otimes$}}

\put(36.03,13.67){\makebox(0,0)[cc]{$\times$}}

\put(12.99,11.71){\dashbox{1.00}(25.00,6.00)[cc]{}}

\put(30.01,40.77){\makebox(0,0)[cc]{x}}

\put(14.03,8.80){\makebox(0,0)[cc]{y}}

\put(40.96,17.77){\makebox(0,0)[cc]{0}}

\put(24.67,3.42){\makebox(0,0)[cc]{d)}}

\put(36.00,8.75){\makebox(0,0)[cc]{SD(II)}}

\put(68.97,13.73){\oval(1.96,2.04)[b]}

\put(66.97,13.73){\oval(2.04,2.04)[t]}

\put(64.97,13.73){\oval(1.96,2.04)[b]}

\put(62.97,13.73){\oval(2.04,2.04)[t]}

\put(60.97,13.73){\oval(1.96,2.04)[b]}

\put(59.00,17.69){\makebox(0,0)[cc]{$q_2^2$}}

\put(69.99,13.71){\line(1,0){21.00}}

\put(90.99,13.71){\line(-4,5){4.00}}

\put(69.99,13.71){\line(3,5){3.00}}

\put(91.03,13.67){\makebox(0,0)[cc]{$\times$}}

\put(68.03,6.80){\makebox(0,0)[cc]{y}}

\put(96.96,12.77){\makebox(0,0)[cc]{0}}

\put(79.67,2.42){\makebox(0,0)[cc]{e)}}

\put(90.00,5.75){\makebox(0,0)[cc]{SD(II)}}

\put(81.03,13.67){\makebox(0,0)[cc]{$\times$}}

\put(24.50,31.50){\oval(9.00,17.00)[]}

\put(24.50,31.50){\oval(13.00,7.00)[]}

\put(18.00,31.00){\line(3,4){2.97}}

\put(19.00,29.00){\line(5,6){4.99}}

\put(26.00,35.00){\line(-3,-4){5.23}}

\put(23.00,28.00){\line(3,4){5.24}}

\put(28.00,28.00){\line(3,4){2.99}}

\put(18.00,30.00){\line(4,5){3.97}}

\put(23.00,35.00){\line(-4,-5){3.98}}

\put(25.00,35.00){\line(-5,-6){5.02}}

\put(22.00,28.00){\line(3,4){5.27}}

\put(24.00,28.00){\line(3,4){5.27}}

\put(30.00,33.00){\line(-4,-5){3.97}}

\put(26.03,27.99){\line(-1,0){1.03}}

\put(25.00,27.99){\line(5,6){5.01}}

\put(31.00,33.00){\line(-4,-5){3.97}}

\put(79.95,41.79){\oval(1.96,1.96)[l]}

\put(79.95,43.79){\oval(1.96,2.04)[r]}

\put(79.95,45.79){\oval(1.96,1.96)[l]}

\put(79.95,47.79){\oval(1.96,2.04)[r]}

\put(79.95,49.79){\oval(1.96,1.96)[l]}

\put(74.02,49.66){\makebox(0,0)[cc]{$q_1^2$}}

\put(79.99,21.71){\makebox(0,0)[cc]{$\otimes$}}

\put(85.01,40.76){\makebox(0,0)[cc]{x}}

\put(79.50,33.41){\oval(13.00,7.00)[]}

\put(73.00,32.91){\line(3,4){2.97}}

\put(74.00,30.91){\line(5,6){4.99}}

\put(81.00,36.91){\line(-3,-4){5.23}}

\put(78.00,29.91){\line(3,4){5.24}}

\put(83.00,29.91){\line(3,4){2.99}}

\put(73.00,31.91){\line(4,5){3.97}}

\put(78.00,36.91){\line(-4,-5){3.98}}

\put(80.00,36.91){\line(-5,-6){5.02}}

\put(77.00,29.91){\line(3,4){5.27}}

\put(79.00,29.91){\line(3,4){5.27}}

\put(85.00,34.91){\line(-4,-5){3.97}}

\put(80.00,29.90){\line(5,6){5.01}}

\put(86.00,34.91){\line(-4,-5){3.97}}

\put(120.97,18.73){\oval(1.96,2.04)[b]}

\put(118.97,18.73){\oval(2.04,2.04)[t]}

\put(116.97,18.73){\oval(1.96,2.04)[b]}

\put(114.97,18.73){\oval(2.04,2.04)[t]}

\put(112.97,18.73){\oval(1.96,2.04)[b]}

\put(111.00,22.69){\makebox(0,0)[cc]{$q_2^2$}}

\put(148.03,10.67){\makebox(0,0)[cc]{$\times$}}

\put(120.03,22.80){\makebox(0,0)[cc]{y}}

\put(151.96,12.77){\makebox(0,0)[cc]{0}}

\put(132.67,2.42){\makebox(0,0)[cc]{f)}}

\put(148.00,5.75){\makebox(0,0)[cc]{SD(II)}}

\put(133.95,43.80){\oval(1.96,1.96)[l]}

\put(133.95,45.80){\oval(1.96,2.04)[r]}

\put(133.95,47.80){\oval(1.96,1.96)[l]}

\put(133.95,49.80){\oval(1.96,2.04)[r]}

\put(133.95,51.80){\oval(1.96,1.96)[l]}

\put(128.02,51.67){\makebox(0,0)[cc]{$q_1^2$}}

\put(133.99,24.71){\makebox(0,0)[cc]{$\otimes$}}

\put(139.01,42.77){\makebox(0,0)[cc]{x}}

\put(133.50,34.50){\oval(9.00,17.00)[]}

\put(133.50,34.50){\oval(13.00,7.00)[]}

\put(127.00,34.00){\line(3,4){2.97}}

\put(128.00,32.00){\line(5,6){4.99}}

\put(135.00,38.00){\line(-3,-4){5.23}}

\put(132.00,31.00){\line(3,4){5.24}}

\put(137.00,31.00){\line(3,4){2.99}}

\put(127.00,33.00){\line(4,5){3.97}}

\put(132.00,38.00){\line(-4,-5){3.98}}

\put(134.00,38.00){\line(-5,-6){5.02}}

\put(131.00,31.00){\line(3,4){5.27}}

\put(133.00,31.00){\line(3,4){5.27}}

\put(139.00,36.00){\line(-4,-5){3.97}}

\put(135.03,30.99){\line(-1,0){1.03}}

\put(134.00,30.99){\line(5,6){5.01}}

\put(140.00,36.00){\line(-4,-5){3.97}}

\put(124.00,15.00){\line(1,0){3.01}}

\put(130.00,15.00){\line(1,0){3.01}}

\put(136.00,15.00){\line(1,0){3.01}}

\put(142.00,15.00){\line(1,0){3.02}}

\put(126.00,22.00){\line(-4,-3){3.97}}

\put(148.00,11.00){\line(-2,3){7.33}}

\put(121.00,10.00){\dashbox{1.00}(28.00,11.00)[cc]{}}

\put(148.00,11.00){\line(-3,-1){11.99}}

\put(123.00,12.00){\makebox(0,0)[cc]{$z_1$}}

\put(146.00,17.00){\makebox(0,0)[cc]{$z_2$}}

\put(122.00,19.00){\line(1,-2){6.00}}

\put(80.00,11.00){\makebox(0,0)[cc]{$z$}}

\put(68.00,9.00){\dashbox{1.00}(25.00,9.00)[cc]{}}

\put(128.00,7.00){\circle*{2.00}}

\put(136.00,7.00){\circle*{2.00}}

\put(79.96,22.96){\line(0,1){2.04}}

\put(79.96,25.98){\line(0,1){2.04}}

\put(79.98,29.98){\line(0,-1){0.95}}

\put(79.50,32.00){\oval(9.03,17.98)[]}

\put(6.00,31.00){\makebox(0,0)[cc]{$+$}}

\put(50.00,31.00){\makebox(0,0)[cc]{$+$}}

\put(105.00,31.00){\makebox(0,0)[cc]{$+$}}

\put(158.00,31.00){\makebox(0,0)[cc]{$+$}}

\end{picture}

\vspace{0.5cm}

\unitlength=2.50pt

\special{em:linewidth 0.4pt}

\linethickness{0.4pt}

\begin{picture}(131.00,53.78)

\put(50.03,13.67){\makebox(0,0)[cc]{$\times$}}

\put(23.03,7.80){\makebox(0,0)[cc]{y}}

\put(54.96,15.77){\makebox(0,0)[cc]{0}}

\put(50.00,7.75){\makebox(0,0)[cc]{SD(II)}}

\put(24.00,14.00){\line(3,-1){9.01}}

\put(50.00,14.00){\line(-3,-1){9.01}}

\put(28.00,20.00){\line(1,0){2.01}}

\put(32.00,20.00){\line(1,0){2.04}}

\put(36.00,20.00){\line(1,0){1.99}}

\put(40.00,20.00){\line(1,0){2.02}}

\put(44.00,20.00){\line(1,0){1.96}}

\put(24.00,14.00){\line(2,3){6.69}}

\put(50.00,14.00){\line(-2,3){6.70}}

\put(36.95,44.80){\oval(1.96,1.96)[l]}

\put(36.95,46.80){\oval(1.96,2.04)[r]}

\put(36.95,48.80){\oval(1.96,1.96)[l]}

\put(36.95,50.80){\oval(1.96,2.04)[r]}

\put(36.95,52.80){\oval(1.96,1.96)[l]}

\put(31.02,52.67){\makebox(0,0)[cc]{$q_1^2$}}

\put(36.99,25.71){\makebox(0,0)[cc]{$\otimes$}}

\put(42.01,43.77){\makebox(0,0)[cc]{x}}

\put(36.50,35.50){\oval(9.00,17.00)[]}

\put(36.50,35.50){\oval(13.00,7.00)[]}

\put(30.00,35.00){\line(3,4){2.97}}

\put(31.00,33.00){\line(5,6){4.99}}

\put(38.00,39.00){\line(-3,-4){5.23}}

\put(35.00,32.00){\line(3,4){5.24}}

\put(40.00,32.00){\line(3,4){2.99}}

\put(30.00,34.00){\line(4,5){3.97}}

\put(35.00,39.00){\line(-4,-5){3.98}}

\put(37.00,39.00){\line(-5,-6){5.02}}

\put(34.00,32.00){\line(3,4){5.27}}

\put(36.00,32.00){\line(3,4){5.27}}

\put(42.00,37.00){\line(-4,-5){3.97}}

\put(38.03,31.99){\line(-1,0){1.03}}

\put(37.00,31.99){\line(5,6){5.01}}

\put(43.00,37.00){\line(-4,-5){3.97}}

\put(22.97,13.73){\oval(1.96,2.04)[b]}

\put(20.97,13.73){\oval(2.04,2.04)[t]}

\put(18.97,13.73){\oval(1.96,2.04)[b]}

\put(16.97,13.73){\oval(2.04,2.04)[t]}

\put(14.97,13.73){\oval(1.96,2.04)[b]}

\put(13.00,17.69){\makebox(0,0)[cc]{$q_2^2$}}

\put(113.03,13.67){\makebox(0,0)[cc]{$\times$}}

\put(86.03,7.80){\makebox(0,0)[cc]{y}}

\put(117.96,15.77){\makebox(0,0)[cc]{0}}

\put(113.00,7.75){\makebox(0,0)[cc]{SD(II)}}

\put(87.00,14.00){\line(3,-1){9.01}}

\put(113.00,14.00){\line(-3,-1){9.01}}

\put(91.00,20.00){\line(1,0){2.01}}

\put(95.00,20.00){\line(1,0){2.04}}

\put(99.00,20.00){\line(1,0){1.99}}

\put(103.00,20.00){\line(1,0){2.02}}

\put(107.00,20.00){\line(1,0){1.96}}

\put(87.00,14.00){\line(2,3){6.69}}

\put(113.00,14.00){\line(-2,3){6.70}}

\put(99.95,44.80){\oval(1.96,1.96)[l]}

\put(99.95,46.80){\oval(1.96,2.04)[r]}

\put(99.95,48.80){\oval(1.96,1.96)[l]}

\put(99.95,50.80){\oval(1.96,2.04)[r]}

\put(99.95,52.80){\oval(1.96,1.96)[l]}

\put(94.02,52.67){\makebox(0,0)[cc]{$q_1^2$}}

\put(99.99,25.71){\makebox(0,0)[cc]{$\otimes$}}

\put(105.01,43.77){\makebox(0,0)[cc]{x}}

\put(99.50,35.50){\oval(9.00,17.00)[]}

\put(85.97,13.73){\oval(1.96,2.04)[b]}

\put(83.97,13.73){\oval(2.04,2.04)[t]}

\put(81.97,13.73){\oval(1.96,2.04)[b]}

\put(79.97,13.73){\oval(2.04,2.04)[t]}

\put(77.97,13.73){\oval(1.96,2.04)[b]}

\put(76.00,17.69){\makebox(0,0)[cc]{$q_2^2$}}

\put(103.50,35.50){\oval(11.00,9.00)[]}

\put(99.00,38.00){\line(-1,-4){0.99}}

\put(99.00,33.00){\line(1,4){1.78}}

\put(100.00,32.00){\line(1,4){2.01}}

\put(103.00,40.00){\line(-1,-5){1.78}}

\put(103.00,31.00){\line(1,6){1.51}}

\put(106.00,40.00){\line(-1,-6){1.49}}

\put(106.00,31.00){\line(1,6){1.32}}

\put(108.00,38.00){\line(-1,-6){1.03}}

\put(108.00,33.00){\line(1,4){0.99}}

\put(22.00,12.00){\dashbox{1.00}(30.00,10.00)[cc]{}}

\put(85.00,12.00){\dashbox{1.00}(29.00,10.00)[cc]{}}

\put(37.00,4.00){\makebox(0,0)[cc]{g)}}

\put(100.00,4.00){\makebox(0,0)[cc]{h)}}

\put(33.00,10.00){\circle*{2.00}}

\put(41.00,10.00){\circle*{2.00}}

\put(96.00,10.00){\circle*{2.00}}

\put(103.00,10.00){\circle*{2.00}}

\put(13.00,33.00){\makebox(0,0)[cc]{$+$}}

\put(67.00,33.00){\makebox(0,0)[cc]{$+$}}

\put(131.00,33.00){\makebox(0,0)[cc]{$-$}}

\end{picture}

\vspace{0.5cm}

\unitlength=2.50pt

\special{em:linewidth 0.4pt}

\linethickness{0.4pt}

\begin{picture}(142.00,46.03)

\put(8.00,28.16){\makebox(0,0)[cc]{$-$}}

\put(133.33,28.16){\makebox(0,0)[cc]{$+\cdots$}}

\put(74.33,28.16){\makebox(0,0)[cc]{$+$}}

\put(36.97,12.98){\oval(1.96,2.04)[b]}

\put(34.97,12.98){\oval(2.04,2.04)[t]}

\put(32.97,12.98){\oval(1.96,2.04)[b]}

\put(30.97,12.98){\oval(2.04,2.04)[t]}

\put(28.97,12.98){\oval(1.96,2.04)[b]}

\put(27.00,16.94){\makebox(0,0)[cc]{$q_2^2$}}

\put(37.99,12.96){\line(1,0){21.00}}

\put(58.99,12.96){\line(-4,5){4.00}}

\put(37.99,12.96){\line(3,5){3.00}}

\put(47.95,37.05){\oval(1.96,1.96)[l]}

\put(47.95,39.05){\oval(1.96,2.04)[r]}

\put(47.95,41.05){\oval(1.96,1.96)[l]}

\put(47.95,43.05){\oval(1.96,2.04)[r]}

\put(47.95,45.05){\oval(1.96,1.96)[l]}

\put(42.02,44.92){\makebox(0,0)[cc]{$q_1^2$}}

\put(47.99,20.96){\makebox(0,0)[cc]{$\otimes$}}

\put(67.01,10.00){\makebox(0,0)[cc]{$p^2$}}

\put(59.03,12.92){\makebox(0,0)[cc]{$\times$}}

\put(35.99,10.96){\dashbox{1.00}(25.00,6.00)[cc]{}}

\put(47.99,28.96){\oval(8.00,14.00)[]}

\put(93.30,12.98){\oval(1.96,2.04)[b]}

\put(91.30,12.98){\oval(2.04,2.04)[t]}

\put(89.30,12.98){\oval(1.96,2.04)[b]}

\put(87.30,12.98){\oval(2.04,2.04)[t]}

\put(85.30,12.98){\oval(1.96,2.04)[b]}

\put(83.33,16.94){\makebox(0,0)[cc]{$q_2^2$}}

\put(94.32,12.96){\line(1,0){21.00}}

\put(115.32,12.96){\line(-4,5){4.00}}

\put(94.32,12.96){\line(3,5){3.00}}

\put(104.28,37.05){\oval(1.96,1.96)[l]}

\put(104.28,39.05){\oval(1.96,2.04)[r]}

\put(104.28,41.05){\oval(1.96,1.96)[l]}

\put(104.28,43.05){\oval(1.96,2.04)[r]}

\put(104.28,45.05){\oval(1.96,1.96)[l]}

\put(98.35,44.92){\makebox(0,0)[cc]{$q_1^2$}}

\put(104.32,20.96){\makebox(0,0)[cc]{$\otimes$}}

\put(125.34,10.00){\makebox(0,0)[cc]{$p^2$}}

\put(115.36,12.92){\makebox(0,0)[cc]{$\times$}}

\put(92.32,10.96){\dashbox{1.00}(25.00,6.00)[cc]{}}

\put(90.32,28.96){\line(1,0){2.00}}

\put(94.32,28.96){\line(1,0){2.00}}

\put(98.32,28.96){\line(1,0){2.00}}

\put(108.32,28.96){\line(1,0){2.00}}

\put(112.32,28.96){\line(1,0){2.00}}

\put(116.32,28.96){\line(1,0){2.00}}

\put(104.32,28.96){\oval(8.00,14.00)[]}

\put(19.00,27.83){\makebox(0,0)[cc]{$\Biggl($}}

\put(142.00,27.83){\makebox(0,0)[cc]{$\Biggr)$}}

\put(48.00,3.00){\makebox(0,0)[cc]{i)}}

\put(105.00,3.00){\makebox(0,0)[cc]{j)}}

\put(56.00,6.00){\makebox(0,0)[cc]{SD(II)}}
\put(80.00,-4.00){\makebox(0,0)[cc]{Fig.1}}

\put(115.00,6.00){\makebox(0,0)[cc]{SD(II)}}

\end{picture}

%



\vspace{.7cm}

Schematically, the appropriate  factorization procedure
 adds  some extra terms into the original OPE.
These terms   are
shown in the next three rows
of Fig.1.
The total contribution
of  terms staying inside  the same bracket
defines the so-called \pagebreak
SD(I)-regime
for every graph of the first row.
In this   regime,  all three currents
are separated by short   distances (cf. \ci{NeRa84b}),
{\it i.e.,} all the intervals $x^2,y^2$ and $(x-y)^2$ are small.
 When  $q_1^2$ goes to zero,
 they have the same singular behavior as the corresponding terms of
the first row. Thus,  in the OPE constructed
for the essentially non-symmetric, small-$q_1^2$  kinematics
\ci{one}, all the  non-analytic terms mentioned above
can be removed, and the final  expression is
regular in the $q_1^2 \to 0$ limit.
The remaining terms  in Fig.1 represent  the
SD(II)-regime, corresponding to a situation when
the electromagnetic current $J_{\mu}(x)$ is  separated by
long  distances from the two other currents,
{\it i.e.}, the interval $y^2$  is small, while  $x^2$ and
$(x-y)^2$  are large.

The short-distance contributions are
factorized into  coefficient functions (CF),
which, in our case (we do not consider radiative corrections),
 are given by a propagator or a product
of propagators. In its turn, the long-distance contribution is represented
by a two-point correlator (see \ci{BalY83,IoSm83,NeRa84b}) of
the electromagnetic current $J_{\mu}(x)$ and
a  composite operator of quark and gluonic fields
 denoted by $\otimes$ in Fig.1.

 We emphasize that it is  the twist\ci{one}
rather than  dimension of
the composite operators which determines the
power behaviour of the contribution
associated with  these bilocals. This  implies that,
for a fixed twist,  one should  include
 the composite operators  with an arbitrary  number $n$ of
derivatives  inside them. 

By definition, the  long-distance factor
(the bilocal)  accumulates nonperturbative information.
One cannot calculate it in a straightforward perturbation theory.
The idea is to incorporate  again the QCD sum rule method.
The starting point is a dispersion relation for the
two-point correlator. 
As the next step,  one should  construct  the OPE for this correlator
in the region of large space-like $q_1^2$ and then
analyze the resulting (``auxiliary'')   QCD  sum rule
to determine the parameters of the relevant model spectral density.
In our case, they  include  the moments
$\langle x^n \rangle_{\rho}$  of the
$\rho^0$-meson wave functions $\varphi_{\rho}(x)$.
To obtain the  correlator at  small $q_1^2$-values,
one should just use this  model spectral density
in  the original dispersion relation
(see also refs. \ci{NeRa84b,BelKog,BeiNeRa88}).

\section{QCD Sum Rule in the
$|q_1^2|\ll|q_2^2|\geq 1{\mbox{ GeV}}^2$ kinematics}

The final result for $\Phi_1(q^2,Q^2,M^2)$ ---
the theoretical part of the sum rule reads\ci{RadRu94}
($q^2\equiv-q_1^2,Q^2\equiv-q_2^2$):
\vspace{0.5cm}

\ba
&\ds \Phi_1(q^2,Q^2,M^2) = \frac{\sqrt{2}\,\al_{e.m.}}{\pi}\,
\frac{1}{M^2} \left\{ \int_0^1 dx \ e^{{-Q^2 x}/{M^2 \bar{x}}}\,
\left\{(1+\frac{q^2x}{M^2}\,e^{{q^2 x}/{M^2}}) \right. \right. +&  \nonumber\\
& &  \nonumber\\
&\ds + e^{{q^2 x}/{M^2}}\,\left[\,\frac{2x}{M^2}\,
\left(q^2\ln{\frac{(s_o+q^2)x}{M^2}} - s_o\right) +
\frac{x^2}{M^4}\,\left(q^4\ln{\frac{(s_o+q^2)x}{M^2}} - q^2 s_o
+\frac{s_o^2}{2}\right)\ \right] - &\nonumber\\
& &  \nonumber\\
&\ds -\left.\sum_{n=1}^{\infty}\,\left(\frac{q^2x}{M^2}\right)^n
\frac{\psi(n)(n+1)}{(n-1)!} \right\} +& \nonumber\\
& &  \nonumber\\
&\ds + \frac{\pi^2}{9}\,
\langle\frac{\al_s}{\pi} GG \rangle\,
\left[\,\frac{1}{2M^2Q^2} + \frac{1}{M^4}\,
\int_0^1 dx\,\frac{x}{{\bar{x}}^2}\,e^{-Q^2x/M^2\bar{x}}\,
\sum_{n=1}^{\infty}\,
\frac{1}{n!}\left(\frac{q^2 x}{M^2}\right)^{n-1}\,\right]
+& \nonumber\\
& &  \nonumber\\
&\ds + \frac{64\pi^3}{243}\,\al_s
{\langle\bar{q}q\rangle}^2\,\frac{q^2}{Q^4M^2}\ + \
\frac{64\pi^3}{27}\,\al_s
{\langle\bar{q}q\rangle}^2\,\frac{1}{2Q^2M^4}\ +& \nonumber\\
& &  \nonumber\\
& &  \nonumber\\
&\ds + \frac{4\pi^2}{3}
\frac{f_{\r}^V m_{\r}}{m_{\r}^2+q^2}\, \int_0^1 dx
\frac{1}{\bar{x}^2 M^2}\, e^{{-Q^2 x}/{M^2 \bar{x}}}\,e^{{q^2 x}/{M^2}}&
\nonumber\\
&\ds \times
\left[\,-a_{V1}\,f_{\r}^V m_{\r}\left(\varphi_{\r_{\bot}}^{V1}(x) -
\frac{4 C_{V51}}{\bar{x} M^2}\,\varphi_{\r_{\bot}}^{V51}(x)\right)
        - f_{\r}^A (1+2\bar{x})\left(\varphi_{\r_{\bot}}^{A}(x) -
\frac{4 C_{A5}}{\bar{x} M^2}\,\varphi_{\r_{\bot}}^{A5}(x)\right)\ \right]&
\nonumber\\
& &  \nonumber\\
& &  \nonumber\\
&\ds +\frac{8\pi^2}{3}\,\frac{f_{\r}^V m_{\r}}{m_{\r}^2+q^2}\,
\int_0^1 d\al\,\al\,\int_0^1 d\bt\,\int_0^1 [dx]_3\
e^{\ b/{a M^2}} &  \nonumber\\
&\ds \times
\left\{f_{3\r}^A\,\varphi_{3\r}^{A}(x_1,x_2;x_3)\,
\left[\,\frac{c_1}{a^2 M^2} - \frac{d_1}{2 a^3 M^4}\,\right] -
  f_{3\r}^V\,\varphi_{3\r}^{V}(x_1,x_2;x_3)\,
\left[\,\frac{c_2}{a^2 M^2} - \frac{d_2}{2 a^3 M^4}\,\right]\right\}&
\nonumber\\
& &  \nonumber\\
& &  \nonumber\\
&\ds - \frac{64\pi^3}{27}
\frac{\al_s \langle\bar{u} u\rangle}{M^4}\,
\frac{m_{\r}\,f_{\r}^V f_{\r}^{T}}{m_{\r}^2 + q^2}
\int\!\!\int_0^1 dx d\bt\,
\frac{\bt\,\varphi_{\r}^{T}(x)}{(1-x\bt)^3}\
e^{\bt(q^2 x\bar{x} - Q^2 x)/((1-x\bt)M^2)}\ -&
\nonumber\\
& &  \nonumber\\
&\ds - \frac{256\pi^3}{27}
\frac{\al_s{\langle\bar{u}u\rangle}^2}{Q^2 M^4}\,
\int\!\!\int_0^1\,d\bt dy\,
\frac{\bt(\bar{\bt}-\bt)y}{(1-y\bt)^3}\
e^{y\bt(q^2\bar{\bt}-Q^2)/((1-y\bt)M^2)}\ -&
\nonumber\\
& &  \nonumber\\
&\ds - \frac{256\pi^3}{27}
\frac{\al_s{\langle\bar{u}u\rangle}}{Q^2 M^4}\,
\frac{m_{\r}\,f_{\r}^V f_{\r}^T}{m_{\r}^2+q^2}\,\frac{3}{4}\,q^2\,
\times & \nonumber\\
& &  \nonumber\\
&\ds \times \left.\int\!\!\int\!\!\int_0^1\,dx d\bt dy\,
\frac{x\bt(\overline{x\bt}-x\bt)y}{(1-x y\bt)^3}\
\varphi_{\r}^T(x)\
e^{y x\bt(q^2\overline{x\bt}-Q^2)/((1-y x\bt)M^2)} \,\right\}  . &
\lab{eq:modifope}
\ea
%
where $f_{\r}^V = 0.2\ \mbox{GeV}, m_{\r} = 0.77\ \mbox{GeV}$ ; the constants
$f_{\r}^A = -f_{\r}^V\,m_{\r}/4,\quad a_{V1} = 1/40$ are  obtained
from the equations of motion\ci{RadRu94}, the values
$f_{3\r}^A=0.6\ \cdot10^{-2}\,\mbox{GeV}^2,
f_{3\r}^V=0.25\ \cdot 10^{-2}\,\mbox{GeV}^2$
are taken from the QCD sum rule estimates given in ref.\ci{CZ84}.

Numerically most important contributions in $\Phi_1$ come from:\
$a)$ SD(I)-regime (first five rows of Eq.\re{eq:modifope}) and
$b)$ $\r^o$-meson contribution with leading twist wave functions
 (SD(II)-regime)
in diagonal and nondiagonal correlators (see also \ci{RadRu94}).

For the continuum threshold in the $\rho$-channel
we take the standard  value\ci{SVZ79} $s_o\simeq 1.5\,\mbox{GeV}^2$
and
asymptotic forms for the following
$\rho$-meson  wave functions \ci{RadRu94,MR86,Rad91}:
\ba
& &\varphi_{V1} = \varphi_{V1}^{as} = 60x\bar{x}\,(2x-1),\quad
\varphi_A = \varphi_A^{as} = 6x\bar{x},\nonumber   \\
& &\varphi_{3A} = \varphi_{3A}^{as} = 360 x_1 x_2 x_3^2,\quad
\varphi_{3V} = \varphi_{3V}^{as} = 7! (x_1-x_2)x_1 x_2 x_3^2 .
\lab{eq:estimas}
\ea
The tensor wave function, however, appears in
a non-diagonal correlator. For this reason,
instead of $\varphi_{T}^{as}=6x\bar{x}$,
we  use
$\varphi_{T}=
\frac{1}{2}( \delta(x) + \delta(\bar{x}) )$ and take
$f_{\r}^T f_{\r}^V m_{\r} = -2 \langle\bar{u}u\rangle$, which
corresponds to the lowest-dimensional contribution
to the nondiagonal correlator (see Fig.1,f).

The terms associated with  three-particle twist-3 wave functions
(Fig.1,e)
are small (contribution of the order  of a few percent).  We expect that
the twist-5 terms  are also small.
We took into account  a specific type of
power corrections, so-called contact terms,
 in a situation when they appear in  the leading-twist bilocals.
However, numerically the contact terms (see Fig.1,h) are  small.

Finally, we write down the  sum rule in the non-symmetric kinematics:
\ba
\fgg(q^2,Q^2)&=&\frac{\sqrt{2}\al_{e.m.}}{\pi f_{\pi}}
\left\{ -2\int_{\s_o}^{\infty} d\s e^{-\s/{M^2}} \int_0^1 dx
\frac{x\bar{x} (q^2 x + Q^2 \bar{x})^2}
{[\s{x}\bar{x} + (q^2 x + Q^2 \bar{x})]^3} \right.\nonumber\\
&+& \left.\Phi_1(q^2,Q^2,M^2)\, \frac{M^2\pi}{\sqrt{2}\,\al_{e.m.}}\right\}.
\lab{eq:finalsr}
\ea

In Fig.2, c) we  plot the
 $\fgg(0,Q^2)$  form factor normalized by the value
$F_{\gamma^*\gamma^*\rightarrow\pi^\circ}^{C.A.}(0,0) =
{\sqrt{2}\al_{e.m.}}/{\pi f_{\pi}}$\ci{ABJ}.
We calculate it in the region
 $Q^2\ge 1 \mbox{GeV}^2$ and compare our
results with  experimental data reported by  CELLO collaboration\ci{CELLO}.
\unitlength1cm
\begin{picture}(5.0,9.5)
\put(1.8,0.3){\epsfxsize12.5cm \epsffile{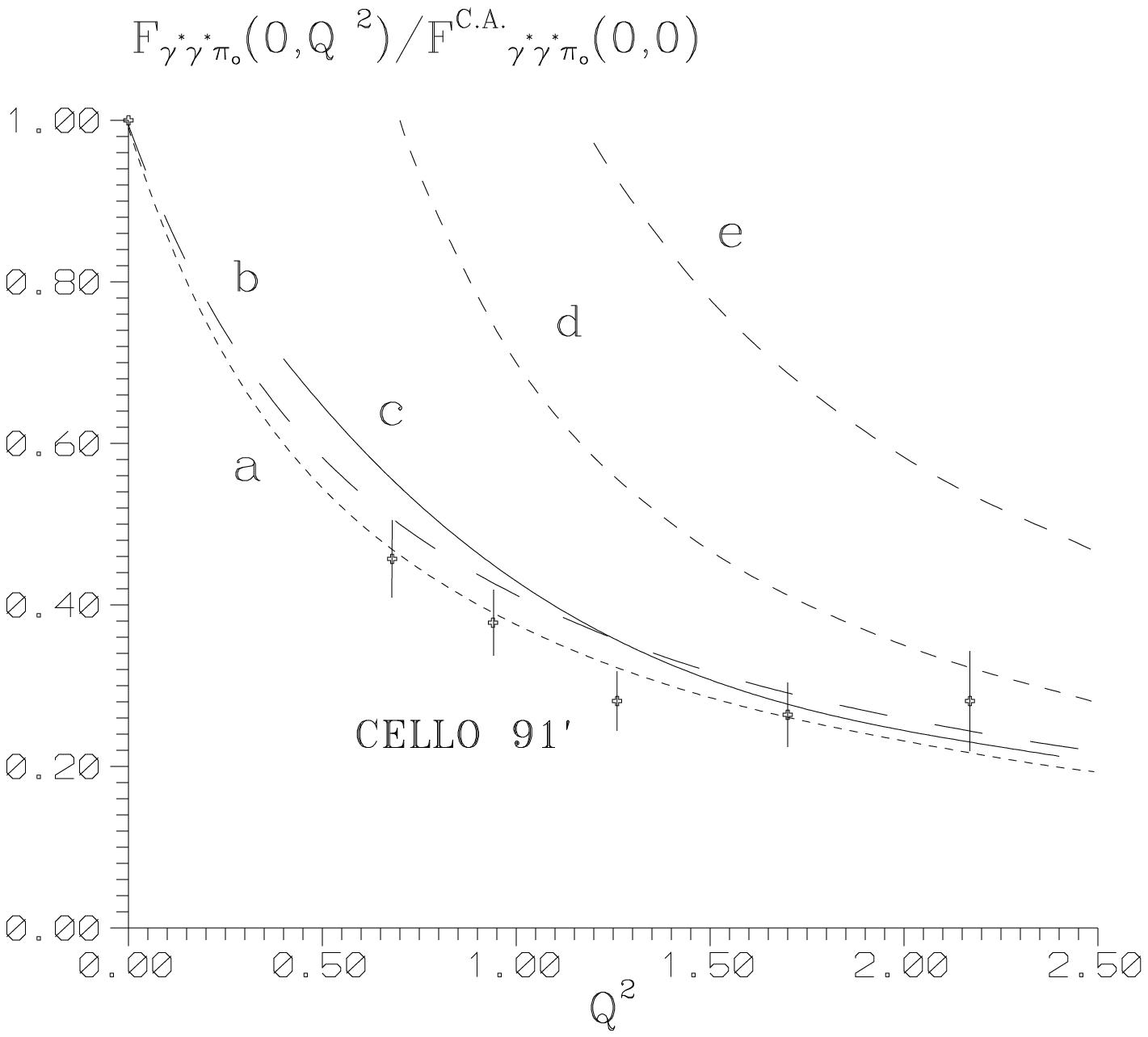}}
\put(6.55,-1.6){\makebox(0,0)[cc]{Fig.2}}
\end{picture}
\vspace{1.8cm}

\noindent
The scale $\s_o$, the continuum
threshold in the pion channel was  obtained by an explicit fitting
procedure. The resulting  values lie
in the interval $0.6\le \s_o \leq 0.85 \, \mbox{GeV}^2$,
{\it i.e.,} they agree with  existing  estimates for the
pion duality interval.
The  sum rule predictions are  rather stable
 in the $M^2$-region
$0.6 \,  \mbox{GeV}^2 \le M^2 \le 1.3 \, \mbox{GeV}^2$ for different $Q^2$.
 Our results agree with experimental data
within an accuracy of
15\% -- 20\%, usual for the QCD sum rules. The other curves prsented in Fig.2
correspond to: a) the vector dominance prediction, b) Brodsky--Lepage
interpolation formula\ci{Brod80}, d) leading twist pQCD-prediction using
asymptotic form for pion wave function, e) leading twist pQCD-prediction using
the Chernyak--Zhitnitsky form for the pion wave function.

Our  sum rule  Eq.\re{eq:finalsr} can be also used
to calculate the form factor
$\fgg(q^2,Q^2)$ at  small (but nonzero) momentum transfer $q^2 \le m_{\r}^2$
and fixed $Q^2\geq 1 \mbox{GeV}^2$.
However, there are no experimental data
for this region. A detailed  analysis of the sum rule
including a detailed study of the sensitivity
to various choices of  the $\r$-meson wave functions
will be given  elsewhere.

The authors are  grateful to A.V.Efremov, S.V.Mikhailov and
A.P.Bakulev for useful  discussions and remarks.
This work was supported in part by Russian Foundation for Fundamental
Research, Grant $N^o$ 93-02-3811,
by International Science Foundation, Grant $N^o$ RFE000
and by US Department of Energy under contract DE-AC05-84ER40150.

\bb{30}

\bi{one} A.V.Radyushkin and R.Ruskov,{\it Phys.At.Nucl.} {\bf 56} (1993) 630.
\bi{RadRu94} A.V.Radyushkin and R.Ruskov, Preprint JINR E2-94-248, Dubna,
1994.
\bi{SVZ79} M.A.Shifman, A.I.Vainshtein and V.I.Zakharov,{\it Nucl.Phys.} {\bf
B147} (1979) 385,448.
\bi{IoSm82} B.L.Ioffe and A.V.Smilga, {\it Phys.Lett.} {\bf B114} (1982) 353.
\bi{NeRa82} V.A.Nesterenko and A.V.Radyushkin, {\it Phys.Lett.} {\bf B115}
(1982) 410.
\bi{NeRa84b} V.A.Nesterenko and A.V.Radyushkin, {\it JETP Lett.} {\bf 39}
(1984) 707.
\bi{BalY83} I.I.Balitsky and A.V.Yung, {\it Phys.Lett.} {\bf B129} (1983) 328.
\bi{IoSm83} B.L.Ioffe and A.V.Smilga, {\it Pis'ma Zh.Eksp.Teor.Fiz.} {\bf 37}
(1983) 250.
\bi{BelKog} V.M.Belyaev and Ya.I.Kogan,{\it Int.J.Mod.Phys.} {\bf A8} (1993)
153.

\bi{BeiNeRa88} V.A.Beilin, V.A.Nesterenko and A.V.Radyushkin, {\it
Int.J.Mod.Phys.} {\bf A3} (1988) 1183.
\bi{CELLO}  CELLO Collaboration H.-J.Behrend et al.,
           {\it Z. Phys.} {\bf C49} (1991) 401.
\bi{CZ84}   V.L.Chernyak and A.R.Zhitnitsky,{\it Phys.Rep.} {\bf 112} (1984)
175.
\bi{MR86} S.V.Mikhailov and A.V.Radyushkin, {\it JETP Lett.} {\bf 43} (1986)
712;\\{\it Sov.J.Nucl.Phys.} {\bf 49} (1989) 494.

\bi{Rad91}  A.V.Radyushkin, {\it Nucl.Phys.} {\bf A527} (1991) 153c.
\bi{Brod80} S.J.Brodsky and G.P.Lepage,{\it  Phys.Rev.} {\bf D22} (1980) 2157;
     {\bf D24}, (1980) 1808.
\nopagebreak
\bi{ABJ} S.L.Adler, {\it Phys.Rev.} {\bf 177} (1969) 2426;
        J.S.Bell and R.Jackiw, {\it Nuovo Cim.} {\bf A60} (1967) 47.
\pagebreak
\eb

\end{document}